\documentclass[twocolumn,twocolappendix]{aastex631}

\newcommand{\spt}{SPT2349$-$56}
\newcommand{\jwst}{{\it JWST}}

\shorttitle{Ultra-red sources in \spt}
\shortauthors{Chapman et al.}

\begin{document}

\title{A large overdensity of LRD-like galaxies discovered by JWST in the SPT2349$-$56 protocluster at $z$=4.30}

\correspondingauthor{Scott Chapman}
\email{scott.chapman@dal.ca}

\author[0000-0002-8487-3153]{Scott C.\ Chapman}
\affiliation{Department of Physics and Atmospheric Science, Dalhousie University, Halifax, NS, B3H 4R2, Canada}
\affiliation{Department of Physics and Astronomy,  University of British Columbia, Vancouver, BC, V6T1Z1, Canada}
\affiliation{NRC Herzberg Astronomy and Astrophysics, 5071 West Saanich Rd, Victoria, BC, V9E 2E7, Canada}
\affiliation{Eureka Scientific Inc, Oakland, CA 94602, USA}

\author[0000-0001-7946-557X]{Kedar A. Phadke}
\affiliation{Department of Astronomy, University of Illinois, 1002 West Green Street, Urbana, IL, 61801, USA}
\affiliation{Center for AstroPhysical Surveys, National Center for Supercomputing Applications, 1205 West Clark Street, Urbana, IL, 61820, USA}
\affiliation{NSF-Simons AI Institute for the Sky (SkAI), 172 E. Chestnut St., Chicago, IL 60611, USA}

\author[0000-0002-3187-1648]{Nikolaus Sulzenauer}
\affiliation{Max-Planck-Institut fur Radioastronomie, Auf dem Hugel 69, Bonn, D-53121, Germany}

\author[0000-0002-6922-469X]{Dazhi Zhou}
\affiliation{Department of Physics and Astronomy,  University of British Columbia, Vancouver, BC, V6T1Z1, Canada}

\author[0000-0003-3596-8794]{Hollis Akins}
\affiliation{Department of Astronomy, The University of Texas at Austin, 2515 Speedway Blvd Stop C1400, Austin, TX 78712, USA}

\author[0000-0002-6290-3198]{Manuel Aravena}
\affiliation{Instituto de Estudios Astrof\'{\i}cos, Facultad de Ingenier\'{\i}a y Ciencias, Universidad Diego Portales, Av. Ej\'ercito 441, Santiago, Chile}
\affiliation{Millenium Nucleus for Galaxies (MINGAL)}

\author[0000-0001-8706-1268]{James R. Burgoyne}
\affiliation{Department of Physics and Astronomy, University of Victoria, Victoria, Canada.}

\author[0000-0002-0930-6466]{Caitlin M. Casey}
\affiliation{Department of Physics,
University of California Santa Barbara, CA, 93106, USA}

\author{Erin Foley}
\affiliation{Department of Physics and Atmospheric Science, Dalhousie University, Halifax, NS, B3H 4R2, Canada}
\author{Leah Fryer}
\affiliation{Department of Physics and Atmospheric Science, Dalhousie University, Halifax, NS, B3H 4R2, Canada}
\author[0009-0008-8718-0644]{Ryley Hill}
\affiliation{Department of Physics and Astronomy,  University of British Columbia, Vancouver, BC, V6T1Z1, Canada}
\author{Nicholas LeVar}
\affiliation{Department of Physics and Astronomy and George P. and Cynthia Woods Mitchell Institute for Fundamental Physics and Astronomy,
Texas A\&M University, 4242 TAMU, College Station, TX 77843-4242, USA}
\author{M\`{o}nica Natalia Isla Llave}
\affiliation{INAF – Osservatorio di Astrofisica e Scienza dello Spazio di Bologna (OAS), Via Gobetti 93/3, I-40129 Bologna, Italy}
\affiliation{Dipartimento di Fisica e Astronomia (DIFA), Universit\`{a} di Bologna, via Gobetti 93/2, I-40129 Bologna, Italy}
\author{Matt Malkan}
\affiliation{University of California, Los Angeles, Department of Physics and Astronomy, 430 Portola Plaza, Los Angeles, CA 90095, USA}
\author{Vismaya Pillai}
\affiliation{Department of Physics and Astronomy,  University of British Columbia, Vancouver, BC, V6T1Z1, Canada}
\author[0000-0002-0741-763X]{William Rasakanya}
\affiliation{Department of Physics, University of Pretoria, Hatfield, Pretoria, 0028, South Africa}
\author{Manuel Solimano}
\affiliation{Centro de Astrobiolog\'ia (CAB), CSIC-INTA, Ctra.\ de Ajalvir km 4, Torrej\'on de Ardoz, E-28850, Madrid, Spain}
\author{Justin S. Spilker}
\affiliation{Department of Physics and Astronomy and George P. and Cynthia Woods Mitchell Institute for Fundamental Physics and Astronomy,
Texas A\&M University, 4242 TAMU, College Station, TX 77843-4242, USA}
\author[0009-0009-5510-8063]{Aurelie Torti}
\affiliation{Instituto de Estudios Astrof\'{\i}cos, Facultad de Ingenier\'{\i}a y Ciencias, Universidad Diego Portales, Av. Ej\'ercito 441, Santiago, Chile}
\author{Joaquin D. Vieira}
\affiliation{Department of Astronomy, University of Illinois, 1002 West Green Street, Urbana, IL, 61801, USA}
\affiliation{Center for AstroPhysical Surveys, National Center for Supercomputing Applications, 1205 West Clark Street, Urbana, IL, 61820, USA}
\author{Fabio Vito}
\affiliation{INAF – Osservatorio di Astrofisica e Scienza dello Spazio di Bologna (OAS), Via Gobetti 93/3, I-40129 Bologna, Italy}
\author{David Vizgan}
\affiliation{Department of Astronomy, University of Illinois, 1002 West Green Street, Urbana, IL, 61801, USA}
\author{George Wang}
\affiliation{Department of Physics and Astronomy,  University of British Columbia, Vancouver, BC, V6T1Z1, Canada}

\begin{abstract}
\noindent
We present 11 newly identified ultra-red galaxies selected with JWST/NIRCam in the core of the $z$=4.3 protocluster \spt. Although these sources are not yet spectroscopically confirmed, their extremely red colours ($m200-m444 > 2.0$ mag), strong spatial concentration, and number density - at least 2000 times that of comparable field populations - make a chance projected association unlikely. Their colours overlap those used to select the widely studied ``little red dot'' (LRD) population, although most of the sources are spatially resolved and are therefore better described as extended red dots (ERDs). The resolved morphologies disfavour continuum emission dominated by an unresolved active nucleus and imply a substantial stellar contribution. Deep ALMA observations put strong limits on obscured star formation, and the ERD population mostly lies significantly below the star-forming main sequence. One object, ERD1/Q1, is substantially brighter and more massive, lying far below the main sequence, making it a strong candidate for a massive quiescent galaxy. 
These JWST-selected red galaxies reveal a 
previously inaccessible population extending  to lower luminosities and stellar masses than the Dusty Star Forming Galaxy (DSFG) population which characterizes the dense protocluster core.
\spt\ may therefore capture an early phase of cluster assembly in which extreme starbursts coexist with galaxies already transitioning toward the passive galaxies characteristic of mature cluster cores.
\end{abstract}



\section{Introduction}


JWST has greatly expanded our understanding of the distant Universe, uncovering galaxies not only at much higher redshifts than ever before, but also entirely new populations of galaxies. One of its most surprising discoveries has been the identification of a new group of compact objects known as ``little red dots'' (LRDs), which appear mainly at $z=4-8$ \citep{Labbe_23a,Akins_23,Matthee_23,akins25}. These LRDs have been found in many of the first JWST surveys by their red color in the longer-wavelength filters of NIRCam, particularly the F277W and F444W bands.  They are also unusually compact in size, making it challenging to separate the galaxy star light from 
any possible AGN at the center. 

Expanding on these LRD studies, \cite{Gentile_24} have recently characterized a complementary population of similarly red but more morphologically extended galaxies in the 0.28~deg$^2$ COSMOS-WEB JWST field, {\it extended red dots} (ERDs). They find that this population is six times less abundant than the LRD population selected in the same field, finding only 61 examples. This is compared to 389 LRDs which satisfy a strict cut in concentration, defined as in \cite{akins25} in the NIRCam F444W band as having C$_{444}>0.5$, where
C$_{444}$=$F_{0.2"}/F_{0.5"}$. While there is still uncertainty about what causes the distinct red color of LRDs (due to light from stars being obscured by dust, or the result of radiation coming from the AGN at their core), in the population of ERDs, the extended envelope of mid-IR light implies that there must be a substantial stellar component in addition to any AGN present.

The clustering of LRDs represents another enigma. The environment of LRDs has been described as {\it lonely} \citep{carranza25}, less dense compared to galaxies at all selections and masses at $z$$>$4. However, close ($\sim$kpc) pairs of LRDs suggest  group-level environments are not uncommon \citep{tanaka24}. It is unclear whether environment plays a large role in the evolution of LRDs. There are no  significant overdensities of LRDs known. 
With the emerging field of high redshift protoclusters enabled by JWST, studying LRDs/ERDs in the densest environments is now possible.

Galaxy clusters grow in the universe through streamed infall and episodic accretion of other large overdensities \citep{springel05, overzier09}.
However, we do not yet know how the early (proto-)cluster environment affects galaxy evolution and what roles may be played by star-formation downsizing \citep{magliocchetti13, miller15, wilkinson17}. 
The existence of dense protocluster cores at $z>4$ like \spt\ 
\citep{miller18}, the DRC \citep{oteo18} strongly suggests that a late-assembly picture of cluster cores may not be fully complete \citep{rennehan20}. Understanding how these massive galaxies form and evolve over time and what regulates and eventually shuts off the star formation in their galaxies is a major  open question in galaxy evolution. 

The massive protocluster core \spt\ (${z\,{=}\,4.3}$) was discovered 
in the 2500~$\mathrm{deg}^2$ South Pole Telescope (SPT) survey \citep{vieira10,everett20} as an incredibly luminous overdensity of star forming, submm-selected galaxies (SMGs) with an integrated S$_{870\mathrm{\mu m}}=110$\,mJy 
corresponding to a SFR $>$$10^{4}~\mathrm{M_\odot} \mathrm{yr}^{-1}$.
The \spt\ intra-cluster medium (ICM) has  been robustly detected through
the thermal Sunyaev-Zeldovich effect (tSZ) using ALMA \citep{zhou2025nat}, ensuring that a substantial hot gas mass is present in the protocluster core.  Evidence for 
different components of the extended ICM gas comes from a large excess of cold gas not obviously connected to the central galaxies \citep{zhou2025apj}. 
 The presence of a significant ICM also implies that ram-pressure processes that help quench SF in dense environments locally can already operate at $z>4$, with a clear example of Jellyfish galaxy identified in \spt\ \citep{zhou2027}. This is essential information for understanding how the passive ``red-sequence'' galaxies that dominate clusters stop forming stars so early. 

An ongoing enigma with \spt\ has been the dramatic overdensity of dusty SMGs (more than 30  --\citealt{sulzenauer25}) contrasted with the relative lack of  non-dusty galaxies selected from the rest-UV \citep{rotermund20,Apostolovski2024}. 
Between the inhospitable environment of the hot ICM, and the overdensity of radio-AGN \citep{Chapman2024,chapman26} which are likely affecting the star formation of galaxies in the core, it is surprising there have not yet been quiescent or quenched galaxies found in \spt. However to date, the photometric depth and requisite red wavelengths from {\it JWST} have not been available to aid in the search, and with {\it Spitzer}-IRAC images being confusion-limited by the massive and dominant SMGs in the \spt\ core \citep{rotermund20,hill2022}.

\begin{figure*}
{\includegraphics[width=17.25cm]{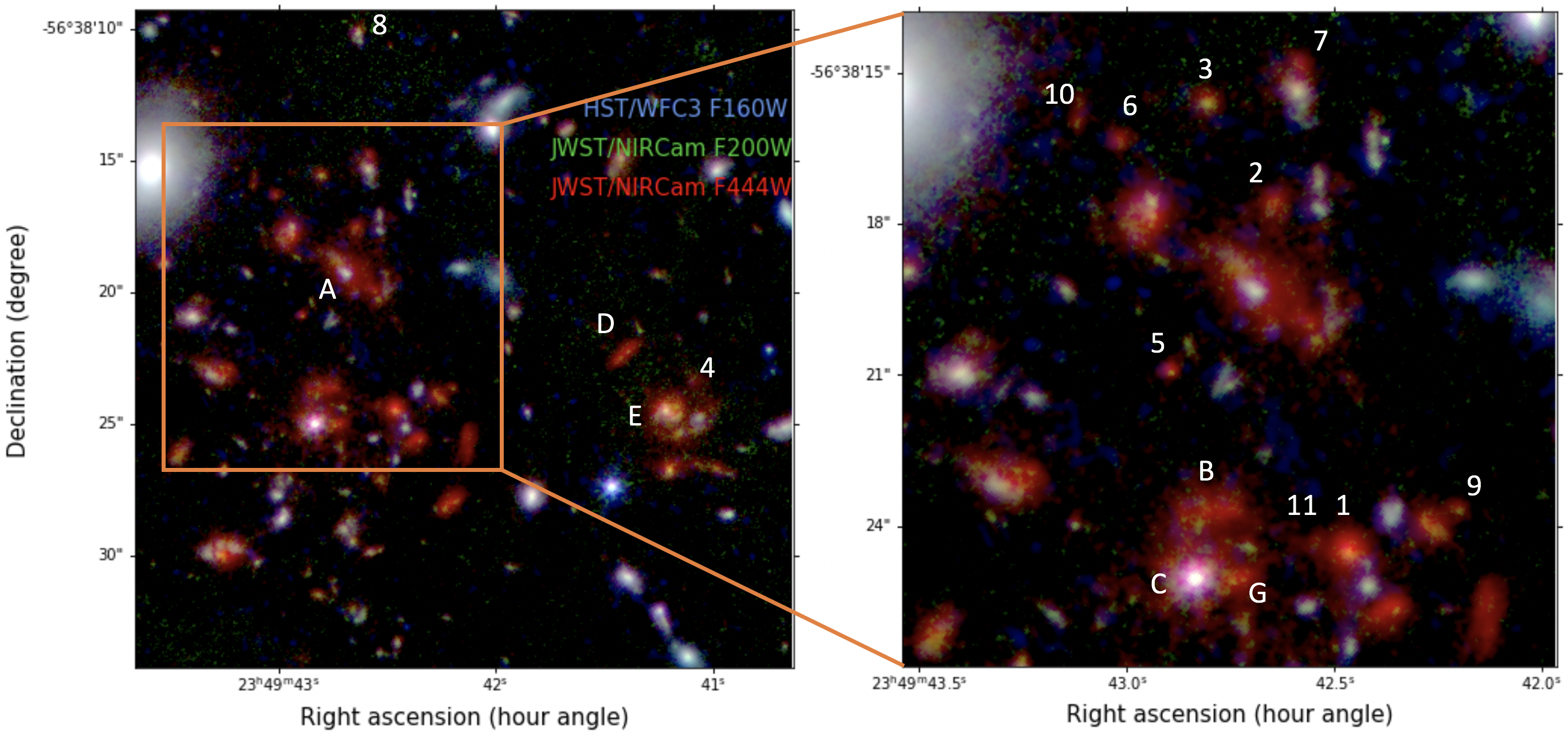}}
{\caption{
{\bf Left:} JWST-based RGB image of the \spt\ protocluster central region (RGB: F444W, F200W, F160W), covering the 11 ERDs discovered. 
The ALMA-selected galaxy members are clearly visible from their generally red colours. {\bf Right:} Zoom of 13$''$ (R=90\,kpc) core region. The brightest of the ALMA-selected SMGs are labelled with letters (from \cite{miller18}) for reference. Radio-AGN have been identified in sources A, C, \& E \citep{chapman26}.
The newly discovered LRD/ERDs are numbered 1--11 (with 4 and 8 lying in the wider field image). A dense subgroup of 6 ERDs lie in a  4$''$$\times$5$''$ region near SMG-A.
}\label{fig:rgb}}
\end{figure*}

In this letter, using \jwst\ NIRCam and MIRI imaging, we set out to find sources in the \spt\ protocluster that were not previously identified in the ultra-deep ALMA continuum and [CII] observations (which have already identified 35 members with [CII]-based redshifts).
 %
 %
Throughout this study, we assume a flat $\Lambda$CDM cosmology with the parameters reported in \citet{Planck_20}, a \citet{Chabrier_03} Initial Mass Function (IMF), and the AB photometric system \citep{Oke_83}.

\begin{figure*}
\centering
{\includegraphics[width=8.1cm]{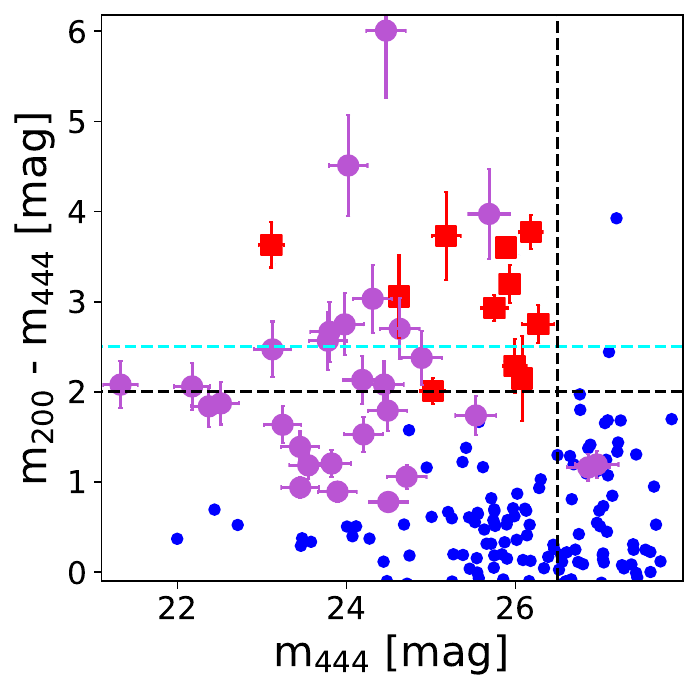}}
{\includegraphics[width=8.1cm]{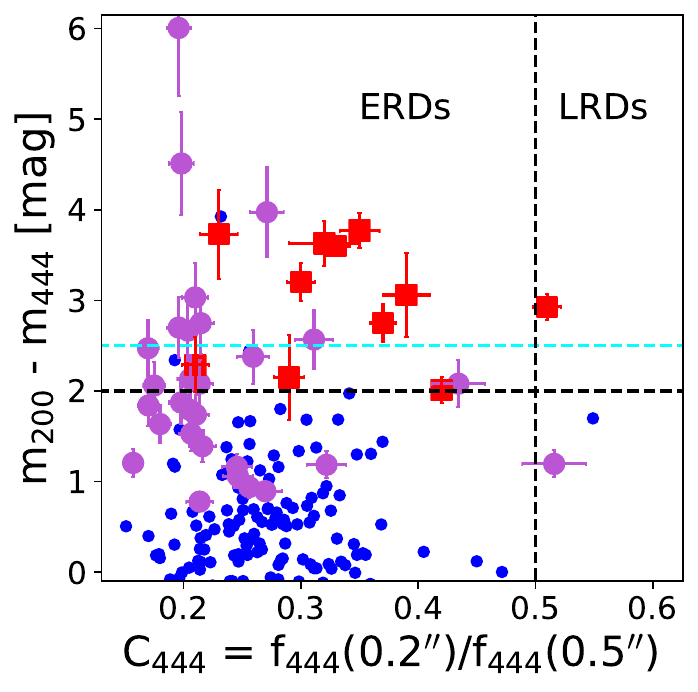}}
{\caption{Selection of LRD/ERDs from F200W-F444W colour (red squares) and comparison to SMGs (orchid circles) in the \spt\ field (remaining sources blue circles). {\bf LEFT:} 
  Color-magnitude diagram in the \spt\ core. SMGs with C+ redshifts in the cluster span a large range in color. We select LRD/ERDs using m$_{200}$-m$_{444}$$>$2.0, which Appendix~\ref{sec:appendixA} shows recovers $>80$\% of comparison LRD/ERD samples with F200W coverage and closely approximates the standard F277W--F444W selection used in the literature \cite{akins25}.
{\bf RIGHT:}  Color-concentration diagram (concentration defined 
C$_{444}$=$F_{0.2"}/F_{0.5"}$), used to select LRDs in the literature \cite{akins25}. One of our red sources (ERD5/LRD1) satisfies the nominal LRD compactness criterion and is unresolved in the F444W and F200W bands.
\label{fig:specplot2}}}
\end{figure*}

\section{Data}
\label{sec:data}
\subsection{JWST Photometry}
\label{sec:JWST_phot}
The  photometry for our sources comes from a followup \jwst\ program on the \spt\ protocluster (GO \#6669, PI Chapman) 
consisting of NIRCam and MIRI imaging.
%
The NIRCam images were taken simultaneously in the F200W and F444W filters with an exposure time of 2920.4\,s. They cover the rest-frame UV and optical emission at $z\sim4.3$. The INTRAMODULE dither pattern was used with four dithers. Following the standard reduction pipeline, the NIRCam images were processed using the \jwst\ Calibration Reference Data System (CRDS) version 11.17.14. The pixel scales of the final images were then resampled to $0.03^{\prime\prime}$/pixel.

The MIRI observations were obtained in the F1000W and F1800W, covering the rest-frame near- and mid-IR. Each filter had an exposure time of 1110.0\,s. The default CYCLING dither pattern with four dithers was used. 
We reduced the MIRI images by primarily following the workflow outlined by the MIRI observation and TEMPLATES teams \citep{rigby_templates_25} using CRDS version 11.18.1. We followed a recommended method from the Space Telescope Science Institute to mitigate the striping artifacts present in all \jwst\ data. We constructed the striping pattern from the median-stacked calibrated images, then subtracted this model pattern from each stage-2 image of the pipeline. This removed the artifacts, though over-subtracted extended emission from larger, nearby sources. However, this is not an issue for this analysis because we focus on the small core region of \spt.
The pixel scale was then resampled to 0.06$''$/pixel. 

To align the MIRI images with NIRCam, we generated a star catalog using \texttt{Source Extractor} on the F444W image to base our alignment on.  Using \texttt{tweakreg} with the F444W catalog, the root-mean-square (RMS) of the final astrometric solutions are 0.01$''$.

\begin{figure}
\centering
{\includegraphics[width=7.5cm]{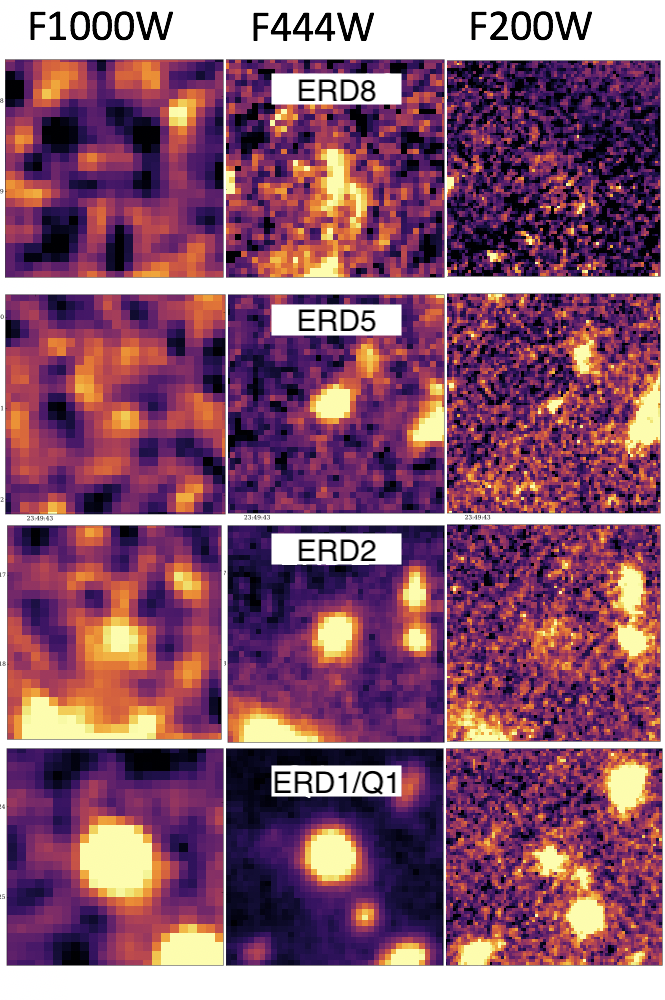}}
{\caption{
 \jwst\ cutouts (3$''$) of 
 example ERDs ranging from the bright ERD1/Q1 (M$^*$$\simeq5\times10^{10}$\,M$_\odot$) through the typical, relatively compact ERD2, the one {\it actual} LRD (ERD5/LRD1), and an extended ERD (ERD8).
}\label{fig:cutouts}}
\end{figure}

\begin{figure}
{\includegraphics[width=8.5cm]{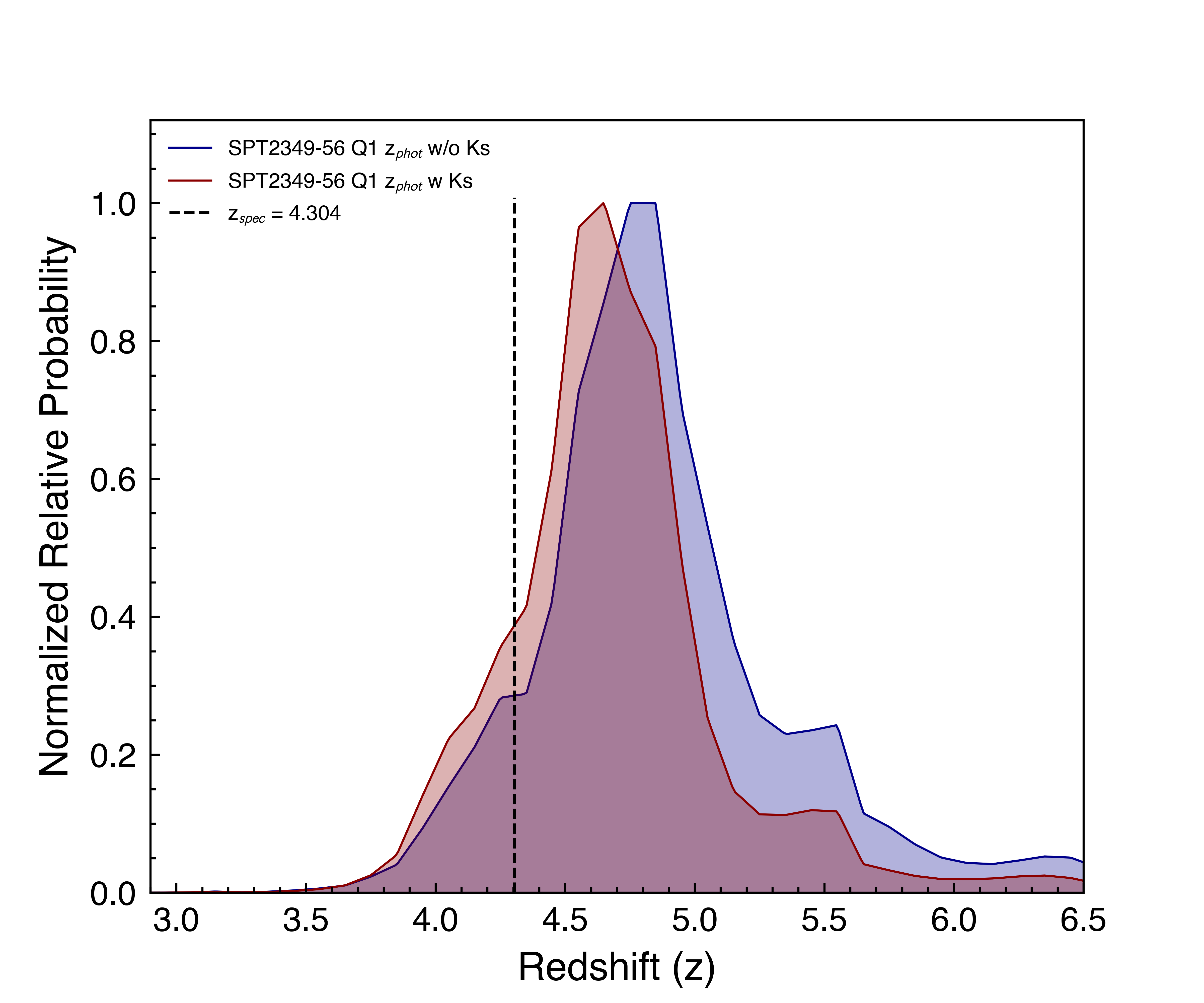}}
{\caption{
Photometric redshift analysis for ERD1/Q1 using \texttt{CIGALE}. 
Although $z$=4.6 is found with highest probability, 
the cluster $z$=4.3 is well within the probability distributions. Note that without using the $K_s$-band detection, the distribution peaks at slightly higher $z$=4.8. The location of the $K_s$ filter entirely longwards of the Balmer-break helps to constrain the photo-$z$ The \texttt{CIGALE} $z$=4.3 SED fit (fig.~\ref{fig:seds}) finds a larger Balmer-break and slightly more reddening than these $z\sim4.6$ solutions.
}\label{fig:q1photoz}}
\end{figure}

\begin{figure*}
\centering
{\includegraphics[width=13.8cm]{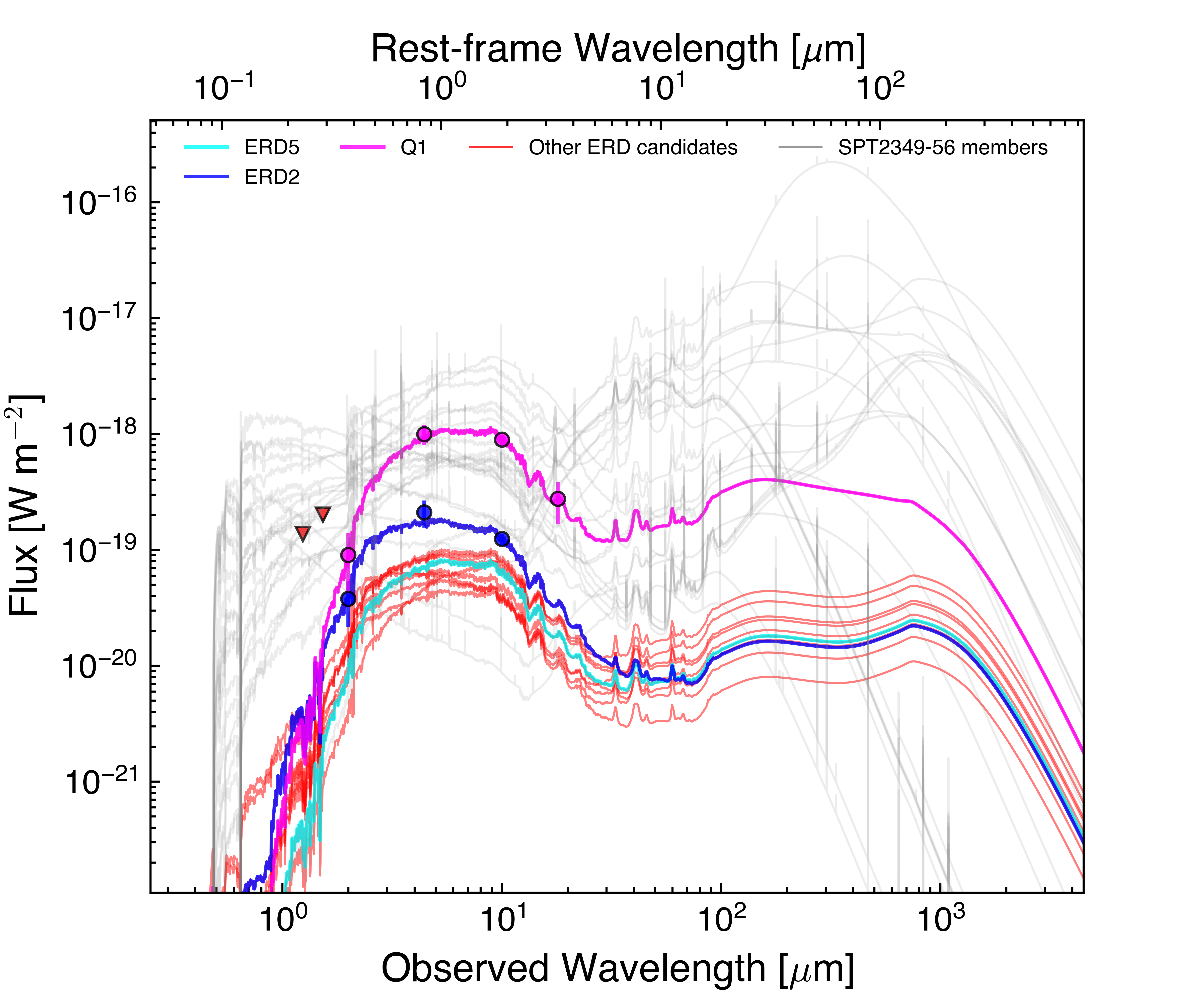}}
{\caption{
SED fits (\texttt{CIGALE}) to the 11 ERD sources identified in the \spt\ core. The sources are generally well fit with old stellar populations and low levels of obscured AGN, but moderate SFRs are permitted by the ALMA constraints.
The ERD-Q1 (consistent with old, quiescent stars) exhibits a stellar mass comparably large to the most massive cluster members previously identified (all high-SFR SMGs -- \cite{rotermund20}). For comparison, all other known cluster members with spec-$z\sim4.31$ are shown grey, mostly exhibiting large submm-wave bumps from their high SFRs.
}\label{fig:seds}}
\end{figure*}

The space-based photometry is extracted with \texttt{sourceXtractor} (\texttt{SE} in the following; \citealt{Bertin_20,Kummel_20}), a model-based tool for extracting photometry from datasets with different spatial resolution. Detection is
conducted on 
the NIRCam F444W filter map. 
By default, \texttt{SE} photometric uncertainties do not include Poisson noise from the background (see e.g., the discussion in \citealt{Akins_23} and \citealt{Casey_23b}). Therefore, we measure the background noise in empty circular apertures and add it in quadrature to the uncertainties estimated by \texttt{SE}, following \citet{Casey_23b}.

\subsection{Red galaxy sample selection}

 Upon visually inspecting the false color RGB maps of \spt\ with NIRCam and {\it HST} imagery (Fig.~1), it is clear that a number of relatively compact (compared to the SMGs) and red galaxies 
 are revealed by \jwst\ in the protocluster core. 
The apparently very red colors and compactness of the new sources seen in the cluster core suggests a comparison to the LRD and ERD populations (e.g. \citealt{Labbe_23a,Labbe_23b,Akins_23,Kokorev_24,PerezGonzalez_24,Gentile_24}), where the densities of these \jwst-selected galaxies have been defined with good statistics over large areas, and over the $z=4-8$ regime in redshift. 
 This motivates a systematic search for possible additional
members of the cluster based on JWST color-selection.

In figure~2, we plot the F444W,F200W colour-magnitude diagram (CMD) of this newly discovered  red compact population, revealing that they share a similar red color distribution with many of the confirmed cluster members, the SMGs and fainter [CII]-emitters most recently cataloged in \cite{sulzenauer25}. Further analysis of the C$_{444}$ compactness criterion of \cite{akins25} reveals the new red sources to generally be more compact than SMGs, although more extended than the LRDs. 
As a starting point we therefore adopt a sample selection consisting of a color cut to collect all the reddest sources in the \spt\ field. To define this color cut, we reference the criterion used in \cite{Barro_23} and analogous to that employed by \cite{akins25}:
\begin{equation}
    F277W-F444W>1.5 \, {\rm mag}
\end{equation}
However, since we only have F200W data in place of F277W, we need to 
extend this cut to shorter wavelengths. Using the average LRD SED from \cite{akins25} implies that the typical LRD in their sample has  
   $ F200W-F444W>2.5 \, {\rm mag}.$
In the case of the \spt\ at $z=4.3$, the F200W filter now spans the Balmer break, possibly creating even redder colors.
In appendix~\ref{sec:appendixA}, we directly assess the LRDs and ERDs in the COSMOS field which are covered by the F200W filter in addition to F277W. This comparison suggests a color cut of    $F200W-F444W>2.0 \, {\rm mag}$  to recover the majority ($>80$\%) of both populations. 
With the compactness criterion of LRDs, the $F200W-F444W>2.0 \, {\rm mag}$ does not introduce any additional galaxies into the sample of LRDs. 
In the case of the ERDs (not requiring the compactness criterion), a 3\% {\it contamination} of additional galaxies is introduced with the bluer color cut. We are thus confident that this $F200W-F444W>2.0$ selection is closely analogous to the $F277W-F444W>1.5$ adopted in the literature.

To reduce the possible contamination by artifacts  and focus only on the robustly detected galaxies, we couple the color cut with 
additional criteria designed to provide at least 2-band detection:

\begin{equation}
    F444W< 26.5 \, {\rm mag}   
\end{equation}
\begin{equation}
    S/N(F200W)>3    
\end{equation}
%
The depth of  the {\it HST} F110W and F160W bands in the \spt\ field \citep{hill2022} 
is substantially shallower than those of other surveys where LRDs have been detected (see e.g. \citealt{Bezanson_22,Finkelstein_23} -- and our Fig.~5). Therefore, we do not expect (nor find) significant emission in these filters even with a similar SED. 
\citet{akins25} have shown how this selection can produce a sample of galaxies with properties analogous to those obtained by adding constraints on the blue shape of the SED (selecting unresolved LRDs in the COSMOS-Web survey with the same color cut employed here).


\begin{deluxetable*}{cccccccccccc}
\label{tab:candidates}
\tablecaption{Observed Properties of the ultra-red galaxies in SPT2349. Sources are ordered (and named) by their $m_{\rm 444}$}
\tablewidth{0pt}
\tablehead{
\colhead{Name} & \colhead{RA} & \colhead{DEC} & \colhead{$m_{\rm 1000}$} & \colhead{$dm_{\rm 1000}$} & \colhead{$m_{\rm 444}$} & \colhead{$dm_{\rm 444}$} & \colhead{$m_{\rm 200}$} & \colhead{$dm_{\rm 200}$} & \colhead{$\Delta m_{\rm 200-444}$} & \colhead{$C_{\rm 444}$} & $\log$M$^*$ \\
\colhead{} & \colhead{(J2000)} & \colhead{(J2000)}  & \colhead{[mag]} & \colhead{[$\delta$mag]} & \colhead{[mag]}& \colhead{[$\delta$mag]}  & \colhead{[mag]} & \colhead{[$\delta$mag]} & \colhead{[$\Delta$mag]} &
 \colhead{f$_{(0.2'')}$/f$_{(0.5'')}$} & \colhead{ [M$_\odot$]}
 }
\startdata
ERD1/Q1 & 23:49:42.463 & -56:38:24.58 & 22.34 & 0.04 & 23.11 & 0.01 & 26.74 & 0.06 & 3.63 & 0.32$\pm$0.05 & 10.9\\
ERD2  & 23:49:42.647 & -56:38:17.67 & 24.66 & 0.20 & 24.62 & 0.03 & 27.68 & 0.15 & 3.06 & 0.39$\pm$0.05 & 9.8\\ 
ERD3  & 23:49:42.809  & -56:38:15.64 & 25.07 & 0.21 & 25.02 & 0.05 & 27.03 & 0.09 & 2.04 & 0.40$\pm$0.05 & 10.2\\ 
ERD4  & 23:49:41.087 & -56:38:23.53 & -- & -- & 25.18 & 0.08 & 28.91 & 0.51 & 3.73 & 0.23$\pm$0.05 & 9.7 \\ 
ERD5/LRD1 & 23:49:42.897 & -56:38:20.98 & -- & -- & 25.75 & 0.04 & 28.68 & 0.15 & 2.93 & 0.52$\pm$0.05 & 9.6\\  
ERD6  & 23:49:43.020 & -56:38:16.41 & 25.38 & 0.23 & 25.89 & 0.06 & 29.49 & 0.44 & 3.60 & 0.32$\pm$0.05 & 9.4\\ 
ERD7  & 23:49:42.586 & -56:38:14.93 & 25.43 & 0.24 & 25.93 & 0.38 & 29.13 & 0.43 & 3.20 & 0.30$\pm$0.05 & 9.0\\ 
ERD8  & 23:49:42.617 & -56:38:08.77 & -- & -- & 25.99 & 0.15 & 28.28 & 0.31 & 2.29 & 0.21$\pm$0.05 & 9.4 \\ 
ERD9  & 23:49:42.208 & -56:38:23.74 & 26.04 & 0.32 & 26.08 & 0.06 & 28.23 & 0.15 & 2.15 & 0.29$\pm$0.05 & 9.6\\ 
ERD10  & 23:49:43.122 & -56:38:16.00 & -- & -- & 26.18 & 0.07 & 29.95 & 0.81 & 3.77 & 0.33$\pm$0.05& 9.1\\ 
ERD11  & 23:49:42.543 & -56:38:24.37 & -- & -- & 26.27 & 0.33 & 29.02 & 0.49 & 2.75 & 0.37$\pm$0.05 & 9.0\\ 
\hline
\enddata 
\tablecomments{
 ERD1/Q1 is additionally detected in F160W with HST (28.21$\pm$0.38 mag),  $K_s$ from Gemini-Flamingos-2 (24.92$\pm$0.17 mag), and F1800W MIRI (23.35$\pm$0.09 mag).}
\end{deluxetable*}
%


These criteria produce a sample of $25$ sources in the $\approx$10 arcmin$^2$ \spt\ NIRCam field, after a first vetting to remove imaging artifacts, all of them lying within the $\approx$0.25 arcmin$^2$ protocluster core. Of these, $11$ are distinct from previously known ALMA-selected submm galaxies.
Our sample includes only one spatially-compact source (ERD5 in Table~1) with C$_{444}=0.52\pm0.11$, lying near the \cite{akins25} C$_{444}>0.5$ interface of LRDs. 
%
Our objects have $F200W-F444W\sim2.9$ on average,  which is actually redder than the nominal LRD selection (with $F277W-F444W>1$ mag or equivalently $F200W-F444W>2$ mag; see e.g.\ \citealt{Labbe_23a,Greene_23,PerezGonzalez_24}), and well above our cutoff color selection. 

We note that by relaxing the selection criteria to fainter magnitudes, we do not admit any obvious additional candidates -- only one further source is seen in the CMD of Fig.~2, and this source is not obviously distinct from the envelope of a much brighter SMG. 
Even relaxing the extreme color-cut does not appreciably add to the sample, with most additional objects down to $F200W-F444W>1.5$ being known SMGs in the cluster.
The sample is thus complete beyond the selection criteria we impose.

\subsection{Ancillary data}
\label{sec:ancillary}
We extract additional photometry for our sources, primarily using the MIRI imaging in this program. We also attempt to  use the other ground/space imaging where it is not heavily confused with neighbouring brighter sources. We extract fluxes in fixed circular apertures from the \jwst-MIRI imaging along with shorter wavelength datasets presented in \citet{hill2022}. These include 
{\it HST}-WFC3 data in the F160W and F110W filters in \cite{hill2022}, and ground based data in \cite{rotermund20} where we use a fixed aperture radius of 1'', well-matched to seeing limits of the ground based-data. 
One of the 11 sources (ERD1/Q1) is detected $>5\sigma$ in both the F1000W and F1800W MIRI bands. Five further sources have $>4\sigma$ detections in F1000W.
None of the 11 sources are detected by HST or ground-based imaging.
The photometry is summarized in Table~1.

\subsection{ALMA data on ERDs}
\label{sec:longwave}

The deep ALMA Band-7 ($\sim850\,\rm \mu m$) continuum data were taken from four different programs in the ALMA Science Archive (Cycle 4: 2016.1.00236.T, PI: S. Chapman; Cycle 5: 2017.1.00273.S PI: S. Chapman; Cycle 6: 2018.1.00058.S, PI: S. Chapman; Cycle 8: 2021.1.01063.S, PI: R. Hill), comprising 25 individual execution blocks and 28 distinct pointing centers. 
We re-ran the full ALMA pipeline in CASA (version 6.6.6-17) to recalibrate all observations in a uniform way. 
The SMGs in the protocluster are very luminous in Band-7 (S/N\,$\gg$\,100), so the imaging is limited by dynamic range rather than thermal noise. 
Low phase errors are required to reduce residuals from the bright sources for recovering nearby faint sources. 
To suppress the phase variations in each measurement set, we performed phase-only self-calibration using the continuum emission from the bright SMGs, which improves the dynamic range and enables deeper deconvolution. 
We used CASA (version 6.7.0.31) to produce the final continuum image with a two-step cleaning strategy. 
First, we flagged all channels affected by strong [CII] emission (357.3\,GHz\,$\sim$\,359.8\,GHz) and imaged the remaining channels with the CASA task {\tt tclean}, using {\tt gridder=`mosaic'}, 
`natural' weighting, 
and the {\tt multiscale} deconvolver, cleaning down to {\tt nsigma=4}. 
The resulting image was used to construct a high-confidence clean mask. 
We selected pixels with values ${\geq}\,4\sigma$ above the local root-mean-square (rms) and then dilated the mask by one synthesized beam to include adjacent pixels with signal ${\geq}\,2\sigma$, so that lower surface-brightness emission can be also included in the clean mask. 
In the second step, we re-imaged the continuum data with the same weighting but using the single-scale {\tt hogbom} deconvolver, further cleaning down to {\tt nsigma=0.5} within this fixed mask to obtain the final continuum map. 
The resulting image reaches an rms of $10\rm\,\mu Jy/beam$ with a synthesized beam of $0.61''\,{\times}\,0.55''$ at full width half maximum (FWHM).

None of the ERDs are individually detected by ALMA. 
We stack the ALMA image at the positions of the 11 ERDs (as described in \citealt{Hill2024}). There is no significant detection in the stack, with only a  $1.7\sigma$ positive signal. The stacked sensitivity is 3$\mu$Jy RMS, for a 3$\sigma$ limit on the average SFR of $<1$\,M$_\odot$\,yr$^{-1}$, assuming a single-component MBB with T$_{\rm dust}$=40\,K at $z=4.3$.
These limits are shown in Fig.~6.

\section{Results}

 Our analysis of the NIRCam imaging shows that many of the red sources fulfill the color selection of widely studied LRDs, but typically have somewhat more extended morphologies with no embedded point source emission (Figs.~1--3). One source ``ERD5'' does nominally satisfy the strict LRD compactness cut. 
 It is notable that while these new sources in \spt\ are mostly not LRDs, they are still dramatically more compact than the SMGs in \spt, which generally appear distinct in Color-C$_{444}$  (Fig.~2) -- SMGs  appear visually {\it enormous} compared to the ERDs in Fig.~1; 
 they are large galaxies even normalized by their stellar mass.
 The ERDs have a median C$_{444}$ of 0.33$\pm$0.12, versus the similarly red SMGs  with median C$_{444}$ of 0.22$\pm$0.11. In fact all SMGs  in \spt\ (including those bluer in 200-444 color) have a similarly large median C$_{444}$ of 0.21$\pm$0.13. 

While none of these objects are yet spectroscopically confirmed, 
 owing to their extremely red colors, along with the small $\sim$300 arcsec$^2$ region in which they are found, they represent an overdensity of $\sim2,000\times$ 
 the ERD field density, found to be 0.06 arcmin$^{-2}$ by \cite{Gentile_24} to a similar depth in F444W.  Five of the brightest and most compact of these ERDs are actually concentrated within a 30~arcsec$^2$ subregion in the north-eastern quadrant of the protocluster core (10,000$\times$ overdense).
 The ERDs are thus highly {\it unlikely} to lie in projection in the protocluster field by chance. 
Including the population of 
14 similarly red SMGs in the protocluster (for a sample of 25 
ultra-red {\it ERD} sources in the same 300 arcsec$^2$ region), an overdensity of $\sim5000\times$ 
the ERD field density is found for the protocluster core as a whole.


\subsection{Photometric Redshifts}


Most of the ERDs in our sample are only detected in 2 or 3 bands (sometimes only robustly in F444W). However, the brighter Q1 is well detected in all four {\it JWST} bands, as well as marginally in the Gemini-F2 K$_s$ band and the HST F160W. It is undetected at all shorter wavelengths with {\it HST} and ground based data.
The strong peak defined by the {\it JWST} imaging allows for a robust photo-$z$ estimate, fully consistent with the $z=4.3$ of the protocluster.  

We derive a photometric redshift for Q1 with 
\texttt{CIGALE} (2025.0) using the stellar models from \citet{Bruzual_03}, {leaving the metallicity as a free parameter in the range $[0.1,1]Z_\odot$,} invoking a delayed exponentially declining star formation history (SFH) with a uniform prior on the e-folding time and on the total stellar mass formed in the ranges [0.3,10] Gyr and [$10^5$,$10^{13}$] M$_\odot$. 
The SEDs are dust-extincted assuming a \citet{Calzetti_00} law with a uniform prior on $A_{\rm v}$ in the range [0,5] mag and a flat redshift prior from [0,10].
The Gemini $K_s$-band lies just above the Balmer-break ($\sim19,300\AA$), and thus potentially has strong diagnostic potential (see Appendix~B). 
However, given the low significance of the $K_s$ detection, we analyze the photo-$z$ with and without $K_s$.

\begin{figure*}
\centering
{\includegraphics[width=12.95cm]{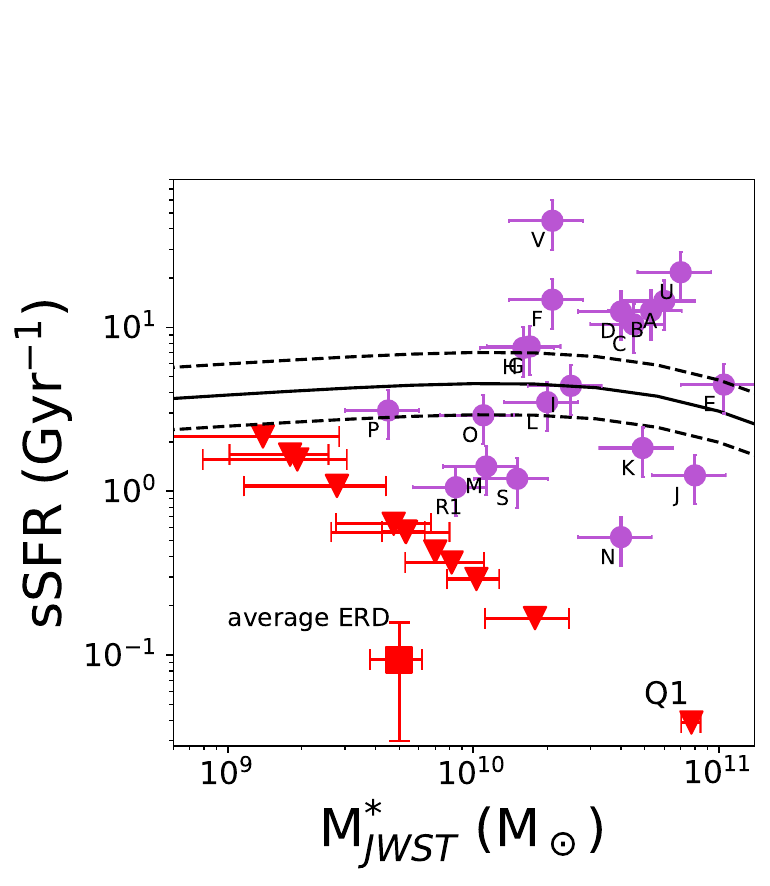}}
{\caption{
 The specific-SFR (sSFR) of the \spt\ ERDs compared with SMGs (orchid circles) (as described in appendix~A). Stellar mass is characterized by CIGALE SED fitting, while SFR is derived from ALMA multi-wave observations, with a 3$\sigma$ limit of 30$\mu$Jy at 850$\mu$m shown for all the ERDs. Red triangles represent those ERDs with at least 3 detected \jwst\ photometric points in the fit, while open triangles depict upper limits in M$^*$ for those with only 2 JWST points. The ERDs range from clearly quenched/quiescent galaxies at the high mass end, down to galaxies consistent with the $z=4$ main sequence of star formation \cite{daddi07} (plotted as $^+_-1\sigma$ envelope).
 The marginal detection from the stacked ALMA 850$\mu$m observation is shown as a red square, suggesting on average the LRDs represent a quenched population.
\label{fig:ssfr}}}
\end{figure*}

As visible in Figure~4, 
the Q1 photo-\textit{z} estimated with and without the K$_s$ band using \texttt{CIGALE} are in good agreement with each other, 
and just slightly higher than the spec-\textit{z} of the protocluster. The K$_s$ version is  peaked at $z=4.6$, while that without K$_s$ peaks at $z=4.8$,
compatible within the uncertainties. 

\subsection{SED-fitting}
\label{sec:sed_fitting}

 Q1 has been reasonably constrained to lie in the protocluster from its photo-$z$. The red NIRCam colours of all the other ERD sources are comparable with Q1, along with the reddest SMGs in the protocluster. 
 Since a photo-$z$ cannot reasonably be constrained for the remaining ERDs (since the  sources are detected in only 2 or 3 bands, and with low SNR in F200W in some cases), 
 we assume they lie in the protocluster (based on the statistical arguments and similarly in colour), and proceed to fit SEDs to all the sources. 
By virtue of the spatial extent of the ERDs (excepting ERD5), their continuum should mostly be attributable to a stellar origin and, therefore, an estimation of the stellar masses less biased by possible AGN (compared with true LRDs) is reasonable .

We derive SED fits with \texttt{CIGALE} \citep{burgarella05,Boquien19}, using the standard set of templates, adopting a redshift of $z=4.30$ for all cases.
\texttt{CIGALE} models a galaxy’s rest-frame optical and ultra-violet spectrum using simple stellar population models with variable star-formation histories, including nebular emission lines, and incorporates a flexible dust attenuation curve that allows the slope and strength of the ultraviolet bump to vary. 
The total current stellar mass, M$^*$, in this model is found by varying the duration of the star-formation, assuming a Chabrier initial mass function (IMF; \citealt{chabrier03}) with solar metallicity. The assumed star-formation history can have an effect on the resulting stellar mass (up to a factor of 2, see \cite{Michalowski12}. Another free parameter is the dust attenuation, given by the amount of extinction present in the V-band in magnitudes, A$_V$, which we model using the \cite{calzetti00}  attenuation curve with a variable power-law slope. 
The SED fitting is shown in Figure~5.

\subsection{Estimated physical properties from SED fits}
\label{sec:properties}

The SED fitting suggests the 11 ultra-red ($m_{200}$-$m_{444}$$>$2.0 mag) galaxies have broadly similar best-fit SEDs (Fig.~5), 
although a majority are constrained by only two or three JWST photometric points. Their red continua can be reproduced by relatively evolved, modestly attenuated stellar populations with little required AGN contribution. None are individually detected in the ultra-deep ALMA Band-7 continuum (S$_{850}<30$\,$\mu$Jy, 3$\sigma$), placing strong limits on obscured star formation (SFR $\lesssim3$\,M$_\odot$ yr$^{-1}$ for the adopted dust model). 

For the well characterized Q1, we find a best fit stellar mass of $\log(M_\star)=(10.9^{+0.07}_{-0.15})$ $M_\odot$ and a dust attenuation of $A_{\rm v}=(0.45^{+0.1}_{-0.1})$ mag, with little dependence on whether an AGN template is introduced (due to the strong suppression of AGN light from the weak 18$\mu$m detection).
Older allowed star-formation histories can raise the inferred mass of Q1 above $10^{11}$\,M$_\odot$, emphasizing the importance of improved spectroscopic constraints on its age and mass-to-light ratio.

For the remaining 10 ERDs, a red continuum with a Balmer break can also explain the SEDs, with stellar masses found in the range $M_\star\sim10^{9.1}$ (ERD9) to $10^{10.2}$\,M$_\odot$ (ERD2), and with little dust extinction (ranging from $A_{\rm v}=(0.23\pm0.19)$ for ERD5 to  $A_{\rm v}=(0.47\pm0.09)$ for ERD6 (see Table~1). 
Balmer breaks imply older stellar populations that have a higher mass-to-light ratio. However a dust obscured AGN can reduce the implied stellar mass in all sources except Q1. By allowing an AGN contribution, the stellar masses are  brought down by factors of two on average, however the constraint is not robust with the poor photometric sampling. We illustrate the effect of different stellar populations and obscured AGN contribution in Appendix~A (Table~2).


Since no galaxy is detected in F110W or F160W, we have only weak limits on the rest-frame UV flux. Hence, the instantaneous SFRs reported by the SED-fitting codes are largely unconstrained; the ALMA limits provide the stronger constraint on obscured star formation. 

We can then compare these sources to SMGs in the \spt\ field in the context of the star forming main sequence (MS) \citep{daddi07} (Fig.~6). 
While the \spt\ ERDs span a large range in M$^*$, the comparison shows that the more massive red objects preferentially lie far below the MS. Q1 is clearly identified robustly as lying far below the main sequence, although approximately half of the ERDs are reasonably ($>5\sigma$) below the MS.
On average (excluding Q1), the population lies $>7\sigma$ below the MS, while the lower-mass sources remain consistent with a broader range of star-formation activity.
With younger, obscured stellar components or reddened AGN contributing to the SED, the stellar masses would be reduced as discussed above, bringing more of them into consistency with the main sequence.

\section{Discussion}
\label{sec:discussion}

JWST observations have revealed an additional galaxy population with extremely red colours in \spt. 
Compared to the SMGs in the protocluster, they have much lower SFRs (constrained by non-detections in ultra-deep ALMA observations). 
SED fitting suggests the red colors are likely from old stellar populations rather than extreme dust reddening or AGN. 
Figure~\ref{fig:ssfr} shows that the ERD population is systematically suppressed relative to the star-forming main sequence, although the lower-mass objects remain consistent with a range of specific SFRs. One galaxy, ERD1/Q1, stands out as being much more massive, and therefore lying the farthest below the main sequence.

\subsection{ERD1/Q1 and the emergence of massive quiescent galaxies in dense environments}
\label{sec:q1}

The most striking individual object in our sample is ERD1/Q1. With a best-fit stellar mass of $M_\star\simeq8\times10^{10}$\,M$_\odot$, modest dust attenuation, and an ALMA limit placing it far below the star-forming main sequence, Q1 is a strong candidate for a massive quiescent or recently quenched galaxy at $z=4.30$. Its photometric-redshift probability distribution is consistent with membership in \spt, although spectroscopy will ultimately be required to confirm the association and establish its star-formation history. If confirmed, Q1 would demonstrate that the transformation of massive galaxies toward the passive population characteristic of mature cluster cores was already underway in this exceptionally dense environment at $z>4$.

An important comparison is provided by \citet{tanaka2024q}, who identified a concentration of massive quiescent galaxies at $z\simeq4$, including a spectroscopically confirmed $M_\star\sim10^{11}$\,M$_\odot$ galaxy. They infer a quiescent fraction of approximately 35\% in the overdensity, compared with only 1.5\% in the surrounding field, and argue that an early red sequence can emerge during the initial collapse of a cluster. The quiescent galaxies in that structure show pronounced Balmer breaks consistent with recent starburst activity followed by rapid quenching. Q1 is comparable in stellar mass to the most massive members of the Tanaka et al.\ quiescent population if fit with similar CIGALE constraints, but it resides in a substantially more extreme and compact protocluster core.

We note that the physical state of \spt\ differs dramatically from the system studied by \citet{tanaka2024q}. Tanaka et al.\ argue that ICM-driven processes are unlikely to dominate in their 
pre-collapse structure, whereas \spt\ already contains a detected hot ICM \citep{zhou2025nat} and direct evidence for ongoing ram-pressure stripping \citep{zhou2027}. At the same time, \spt\ hosts an extraordinary concentration of intensely star-forming SMGs. The coexistence of these extreme starbursts, lower-mass galaxies with suppressed star formation (ERDs), and a massive quiescent candidate (Q1) suggests that \spt\ may be observed during a rapid transition from a starburst-dominated core toward a system containing its first massive passive galaxies. In this sense, Q1 may represent an evolutionary endpoint of the intense star-forming phase. 

The comparison with Tanaka et al.\ also highlights a potentially important mass dependence. Their analysis finds quiescent galaxies predominantly at high stellar masses, while the \spt\ ERD sample extends through $M_\star\sim10^{9.1}$--$10^{10.2}$\,M$_\odot$ and is suppressed relative to the main sequence on average. If spectroscopy confirms that a significant fraction of these lower-mass ERDs are protocluster members with reduced star formation, \spt\ would provide evidence that environmental transformation at $z>4$ extends much farther down the stellar-mass function than has previously been demonstrated.

\subsection{ERDs resulting from environmental quenching}

We first address why we find so many ERDs in \spt. 
A lower-mass and presumably more numerous galaxy population must exist in the protocluster core, but previous observations have been primarily sensitive to the actively star-forming population \citep{sulzenauer25}. A population of lower-mass galaxies with suppressed or quenched star formation could therefore have remained largely undetected before JWST.
Three possible scenarios for the ERDs are outlined below:\\
1.\ {\it External environmental influence from pre-processing.} The recently discovered hot-dense ICM gas in this system \citep{zhou2025nat} may have already had an important transformational effect on the cluster population. 
Indeed one clear example of a `jellyfish' galaxy, revealing ram pressure stripping (RPS) in process, was recently discovered in \spt\ \citep{zhou2027}.
The ERDs may therefore represent a recently quenched star forming population, in some part due to RPS.\\
2.\ {\it AGN feedback from the nearby AGN.} Powerful AGN can have a large impact on surrounding galaxies. Many of the ERDs are  concentrated near the SMG `A/C1', which is the most luminous X-ray AGN in the protocluster \citep{vito24}, with radio jets mapped by ATCA \citep{chapman26}. The ERDs may therefore have very recently been quenched directly from the massive AGN feedback in their surroundings.\\
3.\ {\it Stellar feedback from the ERDs themselves or from nearby SMGs.} Several of the ERDs are close companions with luminous SMGs in \spt, and may have been affected by strong SNe-driven outflows or interactions with their more massive companions. 

All three of these possibilities  represent quenching mechanisms. Because low-mass galaxies are more sensitive to  quenching \citep{geha2012}, this naturally explains why it was difficult to find the signature of environmental influence in previous studies, which concentrated on the more massive protocluster population \citep{hughes25}. 
JWST is therefore essential to identify these quiescent populations and quantify the impact from  overdense protocluster environments in the early Universe.




The spatial juxtaposition of these populations is itself noteworthy. Within the same $\sim$100-kpc-scale core, \spt\ contains extreme dusty starbursts (SMGs), galaxies with modest to strongly suppressed star formation, and the massive quiescent candidate Q1. The protocluster environment therefore does not appear simply to suppress star formation uniformly; instead, it may accelerate galaxy evolution along multiple pathways, triggering intense starbursts in some systems while rapidly exhausting or removing the gas supply in others. \spt\ may thus provide a snapshot of the emergence of the cluster red sequence while the core is still undergoing exceptionally vigorous assembly.

Recent literature results highlight the importance of early quenching.
\cite{kimmig2025}
find that 
galaxies are quenched through a rapid burst of star formation and subsequent active galactic nucleus (AGN) feedback caused by a particularly isotropic collapse of surrounding gas, occurring on timescales of around 200\,Myr or shorter.
\cite{xie2024} show that 
the first quenched massive (M$^*$$\sim10^{11}$ M$_\odot$), Milky Way–mass, and low-mass (M$^*$ $\sim10^{9.5}$ M$_\odot$) galaxies appear respectively at $z\sim4.5$, $z\sim6.2$, and before $z>7$. Most quenched galaxies identified at early redshifts remain quenched for more than 1\,Gyr. In these studies, independent of galaxy stellar mass, the dominant quenching mechanism at high redshift is accretion disk feedback (quasar winds) from a central massive black hole, which is triggered by mergers in massive and Milky Way–mass galaxies and by disk instabilities in low-mass galaxies. Environmental stripping  becomes increasingly more important only at lower redshift.




\subsection{Connection to LRDs and environmental processing of LRD-like galaxies in protoclusters}

Another possibility for the overdensity of ERDs in \spt\ is that they are intimately connected 
to the LRD population described in the literature. 
\cite{billand2026} look at possible LRD descendants and hypothesize a decline in the LRD number density due to the acquisition of stellar components in their outskirts. They propose that as M$^*$ increases the signatures of the LRD fade and the galaxy size grows.
This immense overdensity of ERDs/LRDs in the \spt\ protocluster environment demonstrates that, at least in extreme systems, a strong environmental effect may drive the evolution of the LRD/ERD population, contrasting claims in the literature \citep{carranza25} and further elucidating the mechanisms that might lead to such compact red sources.

In this scenario, the population of compact ERDs in the core of \spt\ may represent environmentally processed analogues of the 
LRD population. 
In the extremely overdense environment of \spt, where dozens of massive galaxies reside within $\sim$130 kpc \citep{sulzenauer25}
the frequency of tidal encounters, minor mergers, and dynamical heating is expected to be substantially higher than in the field,  occurring on short timescales. 
Such processes can efficiently increase galaxy effective radii by redistributing stars to larger orbits while preserving the dense central component (e.g.\ \citealt{naab2009,oser2010}). If LRD-like galaxies are present in the early protocluster population, environmental processing in the core of \spt\ could therefore produce galaxies that retain similarly red colours but exhibit somewhat larger sizes than typical LRDs. In this scenario the compact red galaxies observed in \spt\ represent a transitional phase between the ultra-compact LRD population at high redshift and the dense cores of massive cluster ellipticals that assemble through subsequent merging within the protocluster.


If confirmed spectroscopically, these galaxies may represent the first direct evidence that extreme protocluster environments can rapidly transform ultra-compact high-redshift systems into the dense stellar building blocks of future cluster ellipticals.

More generally, the ERDs may represent only the reddest and most readily identifiable component of a much larger population of lower-mass galaxies in \spt\ that is invisible to existing ALMA-selected surveys. Previous censuses of the protocluster have necessarily been dominated by intensely star-forming systems, whereas JWST is sensitive to galaxies with much weaker obscured star formation. Spectroscopic follow-up will be required to establish how far this population extends to bluer colours and lower masses, where photometric selection becomes increasingly degenerate with foreground and background galaxies.

\section{Conclusions}
\label{sec:conclusions}

We have identified 11 extremely red JWST-selected ERD/LRD galaxies in the \spt\ protocluster field. Their mean colour, $F200W-F444W\simeq2.9$ mag, is substantially redder than our adopted $F200W-F444W>2.0$ selection. Although none is yet spectroscopically confirmed, their concentration within the protocluster core corresponds to an overdensity of at least $\sim2000\times$ relative to the field ERD density, making chance projection of the population highly unlikely. Most are spatially resolved relative to classical LRDs, strongly disfavouring a scenario in which their observed continua are dominated by unresolved reddened nuclear emission and implying a substantial stellar contribution to their red colours.

SED fitting gives typical stellar masses of $M_\star\sim10^{9.1}$--$10^{10.2}$\,M$_\odot$ for the lower-mass ERDs, while the ultra-deep ALMA non-detections place the population systematically below the star-forming main sequence on average. These galaxies therefore reveal a population that was largely inaccessible to previous ALMA- and rest-UV-selected censuses of \spt\ and extend the known protocluster population to substantially lower stellar masses and star-formation rates. The large ERD overdensity, together with their intermediate sizes between classical LRDs and the much more extended SMGs, raises the possibility that at least some are environmentally processed analogues or descendants of LRD-like systems.

Most strikingly, ERD1/Q1 has a best-fit $M_\star\simeq8\times10^{10}$\,M$_\odot$ and lies far below the star-forming main sequence, making it a strong candidate for a massive quiescent or recently quenched member of \spt. The discovery of a concentration of massive quiescent galaxies at $z\simeq4$ by \citet{tanaka2024q} demonstrates that a strong environmental dependence of quenching and an early red sequence can already be present at this epoch. \spt\ provides a complementary and more extreme case: a hot ICM and ongoing ram-pressure stripping are already observed while intense dusty starbursts coexist with suppressed and apparently quenched systems. If Q1 and the lower-mass ERDs are spectroscopically confirmed, \spt\ may capture an unusually early stage in the emergence of the cluster red sequence, when the environmental processes that build the passive population of mature cluster cores are already operating at $z=4.3$.

\section*{acknowledgments}
 
This work is based on observations made with the NASA/ESA/CSA James Webb Space Telescope. The data were obtained from the Mikulski Archive for Space Telescopes at the Space Telescope Science Institute, which is operated by the Association of Universities for Research in Astronomy, Inc., under NASA contract NAS5-03127 for JWST. These observations are associated with program \#6669. Support for US investigators in program \#6669 was provided by NASA through a grant from the Space Telescope Science Institute, which is operated by the Association of Universities for Research in Astronomy, Inc., under NASA contract NAS5-03127.
The National Radio Astronomy Observatory is a facility of the National Science Foundation operated under cooperative agreement by Associated Universities, Inc.
This paper makes use of the following ALMA data: { ADS/JAO.ALMA\#2021.1.01313.S},   ADS/JAO.ALMA\#2018.1.00058.S,  and ADS JAO.ALMA\#2021.1.01010.P. 
ALMA is a partnership of ESO (representing its member states), NSF (USA) and NINS (Japan), together with NRC (Canada), MOST and ASIAA (Taiwan), and KASI (Republic of Korea), in cooperation with the Republic of Chile. The Joint ALMA Observatory is operated by ESO, AUI/NRAO and NAOJ.
We gratefully acknowledge support for this research from NSERC.
This research was supported in part by grant NSF PHY-2309135 to the Kavli Institute for Theoretical Physics (KITP).
Support for this work was provided through grant CSA-GO-01727 awarded by the Canadian Space Agency.
This work was partially supported by the Center for AstroPhysical Surveys (CAPS) at the National Center for Supercomputing Applications (NCSA), University of Illinois Urbana-Champaign.

\bibliography{sample631,sample631_additions}{}
\bibliographystyle{aasjournal}

\appendix

\section{LRDs and ERDs selected with F200W}
\label{sec:appendixA}
\begin{figure}
\centering
{\includegraphics[width=8.95cm]{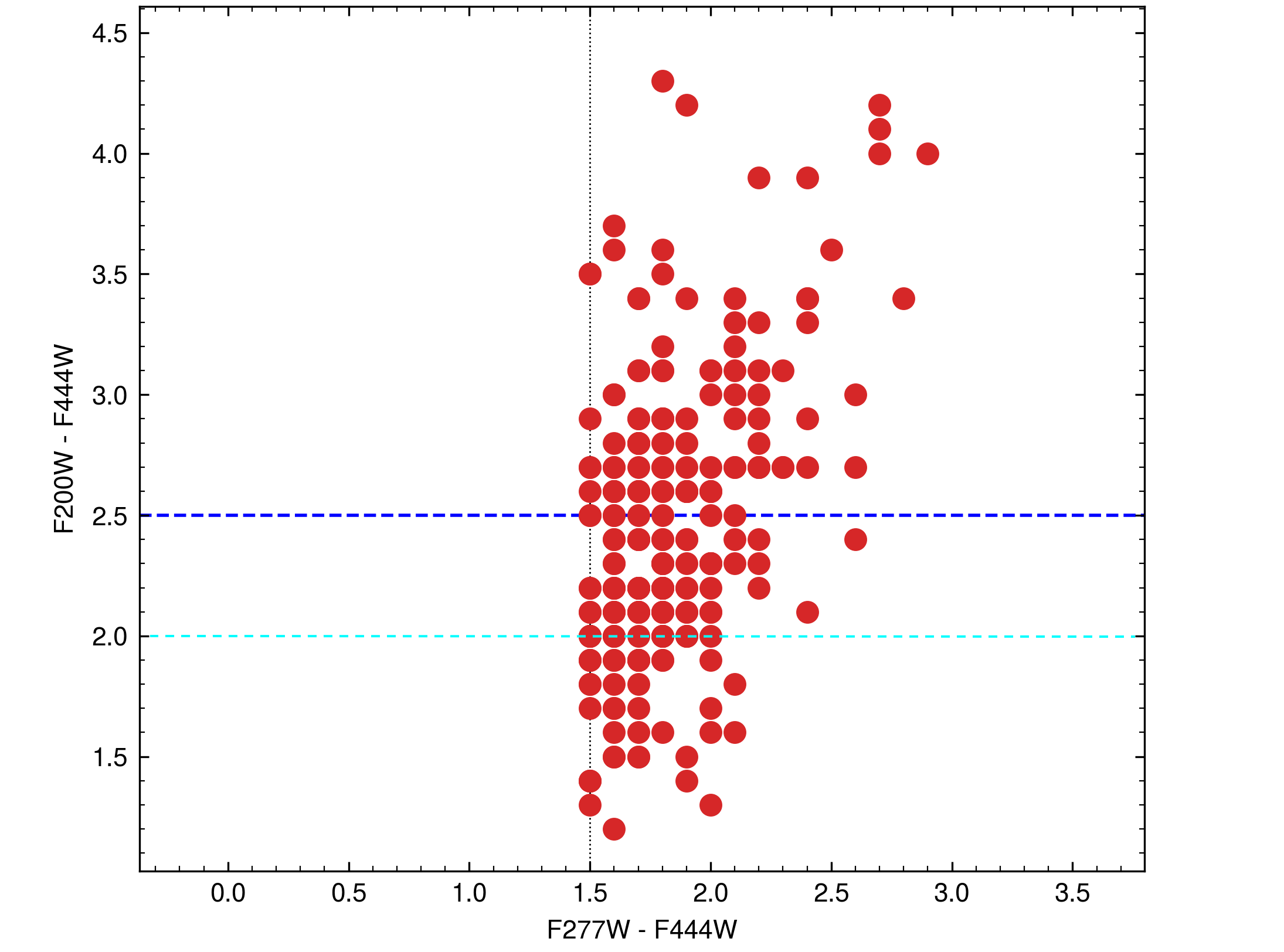}}
{\caption{
 The F200W-F444W vs F277W-F444W distribution for the LRD sample in \cite{akins25}.
\label{fig:lrd200vs277}}}
\end{figure}

 F200W imaging has recently been obtained for a large fraction of the LRDs cataloged in \cite{akins25} and the ERDs cataloged in \cite{gentile24}, by virtue of COSMOS-3D (GO\#5893). There are 224 COSMOS LRDs with F200W coverage, and 105 of those have F200W-F444W $>$ 2.5 (or 47\%). If the selection is reduced down to F200W-F444W$>$2.0, the fraction grows to 184/224 ~or 82\%, which is a clear majority of the population recovered. This advocates to push the selection in \spt\ bluer than the F200W-F444W $>$ 2.5 derived from the average LRD SED fit. This has the virtue of not adding any significant contaminants into the COSMOS samples above, and refining the statistics of LRDs and ERDs for comparison with \spt.
 In the \spt\ sample itself, two additional objects have colours in 2--2.5 range of F200W-F444W. 
Figure~\ref{fig:lrd200vs277} shows the F200W-F444W vs F277W-F444W distribution for the LRD sample in \cite{akins25} with both $>$2.0 and $>$2.5 cuts in F200W-F444W.

To illustrate the degeneracies with our relatively sparse JWST photometry, and to directly connect to the \cite{akins25} criteria, we show  in Table~2 several possible modifications to the {\tt CIGALE} fits for 4 representative ERDs, illustrating that the red colours can manifest in many potential ways in the F277W-F444W filters, whilst retaining the observed F200W-F444W colour selection adopted in this paper.
Moderate to relatively heavy extinction (an average of A$_V$=3.7, assuming the Calzetti dust attenuation law) and a younger stellar population reproduces the observed F200W-F444W colour, but without as strong a Balmer-break, resulting in F277W-F444W colours typically well above 1.5. 
A similar extinction with typical AGN templates are required to reproduce the F200W-F444W colour, again pushing the F277W-F444W colours well above 1.5 (we show both 50\% and 80\% AGN contributions together with the old stellar population. 
AGN contributions with A$_V\sim2$ will actually make the F200W-F444W colours bluer than observed.
If all the ERDs/LRD selected in this paper are truly maximally old stars, then their predicted F277W-F444W colours are on average 1.10 with a dispersion of 0.23. In all 11 ERD cases, either a Bayesian or Best-fit {\tt cigale} solution lies above one in F277W-F444W colour, consistent with literature LRD selection. If somewhat younger stellar populations or AGN contributions are important, then the predicted average F277W-F444W = 1.6$\pm$0.3 for these 11 ERDs.

\begin{table*}
\centering
\caption{
Predicted NIRCam colours for four representative ERDs under several
possible interpretations of their red continua. The ``CIGALE stellar''
rows give the current best-fitting, old stellar-population models.
The ``young dusty'' rows provide an illustrative limiting case in which
a young, approximately flat-$f_\nu$ stellar continuum is reddened with
a Calzetti attenuation law, with $A_V$ chosen to reproduce the observed
$F200W-F444W$ colour. The AGN rows mix the current CIGALE stellar SED
with a reddened AGN continuum, with $f_{\rm AGN,444}$ denoting the
fraction of the total F444W flux contributed by the AGN. 
}
\label{tab:f277_models_simple}

\begin{tabular}{llccccc}
\hline
Source & Model &
$f_{\rm AGN,444}$ &
$A_V$ &
$F200W-F277W$ &
$F277W-F444W$ &
$F200W-F444W$ \\
& & & & \multicolumn{3}{c}{(mag)} \\
\hline

Q1
& CIGALE old stellar & 0.0 & 0.45 & 2.13 & 1.42 & 3.55 \\
& Young dusty stellar & 0.0 & 4.25 & 1.60 & 1.95 & 3.55 \\
& AGN & 0.5 & 2 & 1.31 & 1.29 & 2.60 \\
&     & 0.8 & 2 & 1.06 & 1.22 & 2.27 \\
& AGN & 0.5 & 3 & 1.68 & 1.52 & 3.20 \\
&     & 0.8 & 3 & 1.45 & 1.58 & 3.04 \\
& AGN & 0.5 & 4 & 1.95 & 1.71 & 3.66 \\
&     & 0.8 & 4 & 1.80 & 1.93 & 3.73 \\
\hline

ERD2
& CIGALE old stellar & 0.0 & 0.31 & 1.73 & 1.03 & 2.76 \\
& Young dusty stellar & 0.0 & 3.30 & 1.24 & 1.52 & 2.76 \\
& AGN & 0.5 & 2 & 1.28 & 1.10 & 2.38 \\
&     & 0.8 & 2 & 1.06 & 1.14 & 2.20 \\
& AGN & 0.5 & 3 & 1.56 & 1.29 & 2.85 \\
&     & 0.8 & 3 & 1.42 & 1.48 & 2.90 \\
& AGN & 0.5 & 4 & 1.72 & 1.44 & 3.15 \\
&     & 0.8 & 4 & 1.70 & 1.79 & 3.49 \\
\hline

ERD5/LRD1
& CIGALE old stellar & 0.0 & 0.23 & 1.81 & 1.12 & 2.93 \\
& Young dusty stellar & 0.0 & 3.51 & 1.32 & 1.61 & 2.93 \\
& AGN & 0.5 & 2 & 1.29 & 1.14 & 2.44 \\
&     & 0.8 & 2 & 1.06 & 1.16 & 2.22 \\
& AGN & 0.5 & 3 & 1.59 & 1.35 & 2.93 \\
&     & 0.8 & 3 & 1.43 & 1.51 & 2.94 \\
& AGN & 0.5 & 4 & 1.77 & 1.50 & 3.27 \\
&     & 0.8 & 4 & 1.73 & 1.82 & 3.55 \\
\hline

ERD11
& CIGALE old stellar & 0.0 & 0.33 & 1.60 & 1.14 & 2.74 \\
& Young dusty stellar & 0.0 & 3.28 & 1.24 & 1.50 & 2.74 \\
& AGN & 0.5 & 2 & 1.22 & 1.15 & 2.37 \\
&     & 0.8 & 2 & 1.04 & 1.16 & 2.20 \\
& AGN & 0.5 & 3 & 1.48 & 1.36 & 2.84 \\
&     & 0.8 & 3 & 1.38 & 1.51 & 2.90 \\
& AGN & 0.5 & 4 & 1.62 & 1.52 & 3.14 \\
&     & 0.8 & 4 & 1.65 & 1.83 & 3.48 \\
\hline
\end{tabular}
\vskip0.7cm
\end{table*}
    \label{tab:placeholder}

\section{Updated SED fitting for \spt\ SMGs} \label{sec:appendixB}

In this section we provide {\textsc{CIGALE}} (2025.0) SED fits for \spt\ SMGs 
including new NIRCam and MIRI JWST imaging, updating the fits presented in \cite{Rotermund21,hill2022} which only had  IRAC ch1/2 images available as the longest wavelengths (often highly blended and confused with neighboring galaxies).
There are some clear changes in the stellar masses of several galaxies which were 
largely constrained only by the (often blended) IRAC measurements previously.
We refit these galaxies in {\textsc{CIGALE}}, adding the 4-band \jwst\ photometry presented here. 
For photometry shortward of 1.6$\mu$m, we adopt photometry or limits tabulated in \cite{hill2022}.
{\textsc{CIGALE}} SED fitting presented in \cite{hill2022} is rerun in the same manner with this expanded photometry list.  The resulting M$^*$ constraints 
are used to depict the \spt\ SMGs shown for comparison in Fig.~6. 
More detailed \jwst\ analysis and SED fitting of these galaxies and others in \spt\ from \cite{sulzenauer25} will be presented in an upcoming paper.

Here we briefly outline some of the largest changes from \cite{hill2022}.
Protocluster member N, initially found to have a high M$^*$ in \cite{rotermund20}, was reported as a candidate quenched galaxy. In \cite{hill2022}, using deeper IRAC data and including constraints from HST found a much smaller M$^*$ putting N closer to the main sequence.
With JWST 4-band fluxes, N is now clearly very massive in M$^*$, even larger than reported in \cite{rotermund20}, and is a strong candidate for a galaxy in process of being quenched. 

Galaxy A was essentially unconstrained in \cite{hill2022} due to detection in only one IRAC band. Galaxy~A is now very well sampled with JWST imaging, and the stellar mass is large, but nowhere near the estimate in \cite{hill2022}. Galaxy~A now lies far above the main sequence.
This is true of several other galaxies which  previously had essentially only a single band IRAC detection. In particular, galaxies D, F, N1(U), and N2(V), are all revealed as now being  starbursts lying significantly above the main sequence.

Galaxy C was reported previously as the most massive galaxy in the cluster. V.\ Pillai (in prep.) 
have demonstrated that the bulk of the emission comes from a QSO, and the host galaxy of C is also a starburst above the main sequence. 
Galaxies B and G, with M$^*$ previously completely undetected and unconstrained in \cite{hill2022}, are also highlighted in V.\ Pillai (in prep.). 
They both show  excess above the main sequence ($>2\sigma$ envelope).

\end{document}